\documentclass[12pt]{article}

\title{\bf GAUGE FIELD THEORY\\  FOR POINCAR\'{E}--WEYL GROUP}
\author{\bf O.V. Babourova\footnote{E-mail: baburova@orc.ru},
B.N. Frolov\footnote{E-mail: frolovbn@orc.ru},\\
{\em Moscow State Pedagogical University,}\vspace{-2.5mm}\\
{\em Department of Physics for Nature Sciences,}\vspace{-2.5mm}\\
{\em Krasnoprudnaya 14, Moscow 107140, Russian Federation}\\
{\bf and V.Ch. Zhukovsky\footnote{E-mail: zhukovsk@phys.msu.ru}}\\
{\em Moscow State University, Faculty of Physics,}\vspace{-2.5mm}\\
{\em Department of Theoretical Physics,}\vspace{-2.5mm}\\
{\em Leninskie  Gory, d.1, st. 2, Moscow 119992, Russian Federation}}
\date{}
\begin{document}
\maketitle
\begin {abstract}
On the basis of the general principles of a gauge field theory the gauge theory for the Poincar\'{e}--Weyl group
is constructed. It is shown that tetrads are not true gauge fields, but represent functions from true gauge
fields: Lorentzian, translational and dilatational ones. The equations of gauge fields which sources are an
energy-momentum tensor, orbital and spin momemta, and also a dilatational current of an external field are
obtained. A new direct interaction of the Lorentzian gauge field with the orbital momentum of an external field appears,
which describes some new effects. Geometrical interpretation of the theory is developed and it is shown that as
a result of localization of the Poincar\'{e}--Weyl group spacetime becomes a Weyl--Cartan space. Also the
geometrical interpretation of a dilaton field as a component of the metric tensor of a tangent space in Weyl--Cartan
geometry is proposed.
\end {abstract}

\section {Introduction}
\setcounter {equation} {0}
\par It is well-known that the gauge treatment of physical interactions underlies modern fundamental physics
\cite {Sl-Fad}, \cite {Gauge}. Application of the gauge approach to gravitational interactions was developed in
\cite {Ut}--\cite {BrIvSok} for the Lorentz group and in \cite {Kib1}--\cite {He:rmf} for the Poincar\'{e} group
(see also reviews \cite {Bas}--\cite {Hamm}, the book \cite {Fr:book} and the literature cited there). Nevertheless,
up to present time the interest to the gauge treatment of gravitational interaction \cite {Sard}--\cite {Ald-Sas} does
not weaken. However, the first gauge theory was offered yet in 1918 by Weyl (see \cite {Weyl}), who introduced a gauge
field appropriate to the group of changes of scales (calibres), which were arbitrarily in each point of spacetime.
Change of length scales is equivalent in mathematical sense to expansion or compression (dilatations) of space.
Connection of the dilatation group with the Poincar\'{e} group results in expansion of the Poincar\'{e} group to the
Poincar\'{e}--Weyl group.
\par
Importance of consideration of the Poianc\'{a}re--Weyl group is connected with that role, which the Weyl's scale
symmetry plays in a quantum field theory. Violation of this symmetry at a quantum level results in occurrence of
the Weyl's anomaly connected with the problems of definition of contr-terms structure and asymptotic freedom in the quantum
field theory, with supersymmetry, with calculation of critical dimensions $n=26$ and $n=10$ in a strings theory,
with gravitational instantons, with the Hoking's phenomenon of evaporation of black holes, with the problems of inflation,
the cosmological constant, birthes of particles and black holes in the early universe \cite {Duff}.
For research of some of the listed problems, application of known techniques of BRST-symmetry \cite {BRS}, \cite {T}
to Weyl's gauge scale transformations is used \cite {Boul}.
\par
Construction of a gauge theory for the Poincar\'{e}--Weyl group was developed in \cite {Char-Ta},\cite {Kas}. According to
our opinion, lack of these works consists in the idea (ascending to the work of Kibble
\cite {Kib1}) that the gauge fields for the group of translations are tetrads $h^{a}\!_{\mu}$. This point of view
obviously contradicts the fact that gauge fields should not be transformed as tensors at gauge transformations
while the tetrads are transformed as tensor components on both indexes: tetrad and coordinate ones. We note that
in \cite {Iv-Sar3}, \cite {Sard} it was pointed out on illegitimacy of the treatment of tetrads as gauge fields.
\par In the present work the construction of the gauge theory for the Poincar\'{e}--Weyl group\footnote [1]
{The given extension of the Poincar\'{e} group sometimes is named the Weyl group \cite {Char-Ta}.
From our point of view the name `Poincar\'{e}--Weyl group' $\,$ is more exact, as it is the locale scale
transformation that is usually connected with the concept of the Weyl's symmetry.} is made, deprived the specified lack.
In the basis of this construction there is the method of introduction of gauge fields for the groups connected to
transformations of spacetime coordinates, developed in \cite {Fr1}, \cite {Fr2}, \cite {Fr:book}. The first and the
second Noether theorems are used that allows to introduce the gauge fields, dynamically realizing the appropriate
conservation laws. In the offered approach the quantities $h^{a}\!_{\mu}$ are not gauge fields, but represent some
functions from the true gauge fields. The understanding of what gauge fields are the true potentials of a
gravitational field is obviously important in view of the realization of quantization procedure for the gravitational
field understood as a gauge field for the Poincar\'{e}--Weyl group. Besides, within the framework of general
gauge procedure, the scalar Dirac's field \cite {Dir} and the Utiyama's `measure' scalar field \cite{Uti} are
naturally introduced. These scalar fields play an essential role for construction of a gravitational field Lagrangian.
\par
The paper is organized as follows. In section 2 the Poincar\'{e}--Weyl group and its action on physical fields are
discussed. In section 3 the Noether theorem for the Poincar\'{e}--Weyl group is formulated and there are appropriate laws
of conservation of an energy-momentum, spin and orbital momemta, and also a dilatational current. In section 4,
following \cite {Fr:book}, four initial postulates of the theory are formulated: the principle of local invariance,
the principle of stationary action, the principle of minimality of gauge interactions, and the postulate on existence of
a free gauge field. In section 5 the gauge invariant Lagrangian of the interaction of external and gauge fields is
derived from main principles. A new direct interaction of the gravitational gauge field with the orbital momentum
of an external field appears, which describes some new effects. In section 6 the gauge invariant Lagrangian of free
gauge fields and the gauge fields equations are derived. At last, in section 7 a geometrical interpretation of the
theory is made and it is found out that at localization of the Poincar\'{e}--Weyl group there is a space with a
Weyl--Cartan geometry. In the Conclusion the basic results of the work are discussed, and the resulted Lagrangian of
the gravitational field is proposed,  all gauge fields of the gravitational field gauge theory considered allowing
to be dynamically realized.

\section {The Poincar\'{e}--Weyl group}
\setcounter {equation} {0}
\par Let spacetime $ \cal M $ with a metric tensor $ \breve g $ has structure of a flat space which geometry is
defined according to F. Klein's Erlangen program by the global definition of action of the Poincar\'{e}--Weyl group
$ {\cal PW} (\omega \, \,\varepsilon \, \,a) $ as a fundamental group (on E. Cartan's terminology \cite {Cart}) of
this geometry. The fundamental group determines geometry of space as system invariants, that is relations
and geometrical images, which remain constant in space at action of the given group (see \cite {Sh-Str}).
The similar geometry arises in spacetime filled with radiation and ultra relativistic particles. The space is flat,
when curvature, torsion and nonmetricity tensors are equal to zero at all space. Such type of space $ \cal M $
is appropriately to name as Minkowski--Weyl space.
\par
On $ \cal M $ a special system of coordinates $x ^ {i} $ ($i =1,2,3,4 $) (analogue of the Cartesian coordinates
in Minkowski space) can be globally introduced, in which metric tensor components are equal
\begin {equation} g _ {ij} = \beta ^ {2} g ^ {M} _ {ij} \;, \quad \beta = const > 0
\;, \quad g ^ {M} _ {ij} = \mbox {diag} \, (1, \, 1, \, 1, \,-1) \;, \label{eq:gM1}
\end {equation}
where $g ^{M} _ {ij} $ are components of the metric tensor of Minkowski space.
\par
We represent infinitesimal transformations of the group ${\cal PW}$ as follows:
\begin {eqnarray}
&&\delta x ^ {i} = \omega^m I_m {} ^ {i} \! _ {j} \, x ^ {j} - \varepsilon x^i + a^i =
- (\omega^m \stackrel {\circ} {M} _ {m} + \; \varepsilon\stackrel {\circ} {D} + \; a^k P_k) x ^ {i} \;,
\label {eq:1004} \\
&&\stackrel {\circ} {M} _ {m} = - I_m {} ^l \! _ j \, x ^ {j} \frac{\partial} {\partial x ^ {l}} \;,
\quad I_m {} ^ {ij} = I_m {} ^ {[ij]} \;, \quad \stackrel {\circ} {D} = x^l \frac {\partial} {\partial x ^ {l}} \;,
\quad P _ {k} = - \, \delta ^ {l} _ {k} \frac {\partial} {\partial x ^{l}} \;. \label {eq:104}
\end {eqnarray}
Introducing the generalized designation $\{ \omega^{z} \} = \{ \omega^{m}\,, \, \varepsilon\,,\, a^k \}$
for a set of parameters of trans\-for\-ma\-ti\-ons of the given group, the transformation (\ref {eq:1004}) is
convenient to represent as
\begin{equation}
\delta x^{i} = \omega^{z}X^{i}_{z}\; , \quad X^{i}_{m} = I_{m}{}^{i}\!_{j}\, x^{j} \; , \quad
X^i = -\,x^i \; , \quad X^{i}_{k} = \delta^{i}_{k}\; . \label{eq:1006a}
\end{equation}
where $I _ {\mbox{\o}} {} ^ {i} \! _ {j} = - \,\delta_j \! ^ i $, and $ \mbox{\o} $ is a symbol of empty set.
\par
Operators $ \stackrel {\circ} {M} _ {m} $ and $P _ {k} $ are generators of 4-rotations (Lorentz subgroup $L_4$)
and 4-shifts (a subgroup of translations $T_4 $), and the operator $ \stackrel {\circ} {D} $ is a generator
of dilatation (a subgroup of dilatation $D_4 $) of the space $ \cal M $. The given operators satisfy to the following
commutation relations:
\begin{eqnarray}
&&\left [\stackrel{\circ}{M}_{m}\,,\,\stackrel{\circ}{M}_{n}\right ] = \;
c_m\!^q\!_n \stackrel{\circ}{M}_{q}\;, \qquad \left [\stackrel{\circ}{M}_{m}\,,\,
P_{k}\right ] =I_m{}^l\!_k P_{l}\;,\qquad \left [P_{k}\,,\,P_{l}\right ]= 0\;,\nonumber\\
&&\left [\stackrel{\circ}{M}_{m}\,,\,\stackrel{\circ}{D}\right ] = 0\;,\qquad
\left [P_{k}\,,\,\stackrel{\circ}{D}\right ] = P_k\;, \qquad
\left [\stackrel{\circ}{D}\,,\,\stackrel{\circ}{D}\right ] = 0\;. \label{eq:CR}
\end{eqnarray}
\par
Let an arbitrary field $ \psi ^ {A} $ is given on $ \cal M $, the infinitesimal transformation of which under
action of the group $ {\cal PW} (\omega , \,\varepsilon , \, a) $ looks like
\begin{equation}
\delta \psi^A =\omega^{m}I_{m}\!^{A}\!_{B}\psi^{B} + \varepsilon w\psi^A =
\omega^z I_z\!^{A}\!_{B}\psi^{B}\;, \quad I_z\!^{A}\!_{B} = \{ I_m\!^{A}\!_{B}\,,\,
I_{\mbox{\o}}\!^{A}\!_{B}\}\;,\quad  I_{\mbox{\o}}\!^{A}\!_{B} = w\delta^A_B \;, \label{eq:1022}
\end{equation}
where $w$ is a weight of $\psi^A$ under the action of the subgroup of dilatation $D_4$.
Operators $I_m \! ^ {A} \! _ {B} $ satisfy to commutation relations:
${I}_{m}\!^{A}\!_C\,I_{n}\!^{C}\!_B - I_{n}\!^{A}\!_C\, I_{m}\!^{C}\!_B =
c_{m}\!^{q}\!_n\, I_{q}\!^{A}\!_B$.
\par
Action of group $ {\cal PW} $ on metric tensor occurs as follows:
\begin {equation}
\delta g _ {ij} = - \, \omega ^ {z} \, I _ {z} {} ^ {l} \! _ {i} \, g _ {lj} - \omega^z \, I_z {} ^ {l} \! _ {j}
\, g _ {il} = - \, 2\omega^z \, I _ {z (ij)} = 2\varepsilon \, g _ {ij} \;. \label {eq:vg}
\end {equation}
Components $g _ {ij} $ are not invariant under the action of ${\cal PW}$. As a result of (\ref{eq:gM1}),
under the action of ${\cal PW}$ the following transformation holds,
$\beta^2 \longrightarrow \beta^2 + 2\varepsilon\,,\;\;\delta \beta = \varepsilon \beta = \omega^{z}\delta^{\mbox{\o}}_z\,\beta$.
\par
We introduce on $ \cal M $ an arbitrary curvilinear system of coordinates $\{x^{\mu}\}= \{x^{\mu}(x^{i})\}$:
\begin{equation}
dx^{i} =\stackrel{\circ}{h}\!^{i}\!_{\mu} dx^{\mu}\;, \qquad\stackrel{\circ} {h}\!^{i}\!_{\mu} =
\partial_\mu x^i \; , \qquad  \partial_\mu = \frac{\partial}{\partial x^{\mu}}\;. \label{eq:dx}
\end{equation}
Then for metric tensor it is valid:
\begin{eqnarray}
&& ds^{2} = \stackrel{\circ}{g}_{\mu\nu} dx^{\mu} dx^{\nu}\;, \qquad
\stackrel{\circ}{g}_{\mu\nu} =  \breve g(\vec{e}_{\mu}\,,\,\vec{e}_{\nu}) =
g_{ij} \stackrel{\circ}{h}\!^{i}\!_{\mu}\stackrel{\circ}{h}\!^{j}\!_{\nu} \; ,\nonumber\\
&& \stackrel{\circ}{g} =\mbox{det}(\stackrel{\circ}{g}_{\mu\nu}) = \mbox{det}(g_{ij})\,(\stackrel{\circ}{h})^2\;,
\quad \stackrel{\circ}{h} = \mbox{det}(\stackrel{\circ}{h}\!^{a}\!_{\mu} )\;.  \label{eq:deth}
\end{eqnarray}
The curvilinear system of coordinates in flat space is introduced with the purpose to separate the
problems connected with invariance of the theory under gauge transformations, and the problems following from the
requirement of covariance of the theory concerning the group of the general transformations of coordinates.
In works \cite {Kib1}--\cite {Fr2}, \cite {Char-Ta}, \cite {Kas}, the curvilinear
system of coordinates was not introduced before procedure of localization. As a result, transformations of
coordinates under action of gauge groups after localization became the general transformations of
coordinates that broke mathematical structure of gauge groups.
\par
As a result of (\ref {eq:1022}) and (\ref{eq:vg}), both $\psi^A (x)$ and $g_{ij}$ are not transformed as representation
of the subgroup of translations $T_4$, but change under action of $T_4$ only owing to trans\-for\-ma\-ti\-on of argument
$x^k$. Therefore an action of the operator of shift on ${\cal M}$, for example, on a field $\psi^A(x)$
(when $\delta x^{k} = a^{k}$) is realized as follows:
\begin{eqnarray}
&&\psi^{A} (x + \delta x) = \psi^{A}(x) + \delta x^{\mu}\frac{\partial}{\partial
x^{\mu}}\psi(x) = \psi^{A}(x) - a^k P_k\psi^A (x) \; , \label{eq:1014a}
\\ && P_k = - \stackrel{\circ}{h}\!^{\mu}\!_k \,\partial_\mu \;,
\qquad  \stackrel{\circ}{h}\!^{\mu}\!_k \stackrel{\circ}{h}\!^k
\!_{\nu} = \delta^{\mu}_{\nu}\;. \label{eq:1014}
\end{eqnarray}
\par
Under the action of ${\cal PW}(\omega\,,\,\varepsilon\,,\,a)$, variation $\bar{\delta}$ of a form of
the field function $\psi^A$:
\begin{eqnarray}
&&\bar{\delta} \psi^A = \psi'^A (x) - \psi^A (x) =\delta \psi^A -\delta x^\mu\partial_\mu
\psi^A = \delta \psi^A + \delta x^k P_k \psi^A = \nonumber \\
&& = \delta \psi^A + \omega^z X^k_z P_k \psi^A = \delta \psi^A - \omega^z
X^k_z \stackrel{\circ}{h}\!^{\mu}\!_k\partial_\mu \psi^A \;, \label{eq:dbar}
\end{eqnarray}
commutes with the operator of differentiation. The following commutation relations hold:
\begin{equation}
[\bar\delta\,,\,P_k ] = 0\;, \qquad
[\delta\,,\,P_k ]= (P_k \delta x^l)\,P_l \;,\label{eq:com}
\end{equation}
which are valid owing to identities: $\;\delta \stackrel{\circ}{h}\!^{\mu}\!_k =
-\stackrel{\circ}{h}\!^{\mu}\!_l \stackrel{\circ}{h}\!^{\nu}\!_k \,\delta
\stackrel{\circ}{h}\!^l\!_{\nu}\;$ and
\[
\bar\delta \stackrel{\circ}{h}\!^l\!_{\nu} = \partial_\nu \delta x^l - \stackrel{\circ}
{h}\!^l\!_{\mu} \partial_\nu\delta x^{\mu} - \delta x^\mu \partial_\mu\!\stackrel{\circ}{h}\!^l\!_{\nu}
= 2\partial_{[\nu}\!\stackrel{\circ}{h}\!^l\!_{\mu ]}\,\delta x^{\mu} = 0 \;.
\]

\section {Noether theorem for the Poincar\'{e}--Weyl group}
\setcounter {equation} {0}
\par
The field $\psi^A$ on $\cal M$ is described in curvilinear coordinates by the action
\begin{equation}
J = \int_\Omega \,(dx)\,{\cal L}\,,\qquad {\cal L} = \;\sqrt{\mid \stackrel{\circ}{g}\mid}\,
L(\psi^A,\,P_{k}\psi^A,\,\beta^{2} g^{M}_{ij})\;. \label{eq:1017}
\end{equation}
Under transformations of $\cal M$ under the action of the Poincar\'{e}--Weyl group (\ref {eq:1004}),
variation of action integral with regard of change of integration area $\Omega $ is equal
\begin{equation}
\delta J = \delta \int_\Omega \,(dx)\,{\cal L} = \int_\Omega \,(dx)\Biggl (\sqrt{\mid \stackrel{\circ}{g}\mid}\;
(\partial_\mu \delta x^\mu )\, L + \delta \sqrt{\mid \stackrel{\circ}{g}\mid}\; L +
\sqrt{\mid \stackrel{\circ}{g}\mid}\;\delta L \Biggr ) =0 \;.\label{eq:dJ0}
\end{equation}
\par
In curvilinear system of coordinates, as consequence of (\ref {eq:deth}) an equality holds,
$\delta \sqrt{\mid \stackrel{\circ}{g}\mid}\, = - \sqrt{\mid \stackrel{\circ}{g}\mid}\,(\partial_\mu \delta x^\mu )$.
Therefore (\ref {eq:dJ0}) by virtue of randomness of area $\Omega$ means $\delta L = 0$:
\[
\delta L = \frac{\partial L}{\partial \psi^A}{\delta}\,\psi^A + \frac{\partial L}{\partial P_k \psi^A}
\,{\delta}P_k \psi^A + \frac{\partial L}{\partial \beta}\,\delta \beta = 0\; .
\]
Calculating the variations on the basis of (\ref {eq:1022})--(\ref {eq:com}) and using equality
$P_k X^i_z = - I_z {} ^i \! _ k $, we receive as consequence of randomness of $\omega^z$ the following identity:
\begin{equation}
\frac{\partial L}{\partial \psi^A} I_{z}\!^{A}\!_{B} \psi^B + \frac{\partial L}{\partial P_k
\psi^A} (I_{z}\!^{A}\!_{B} P_k \psi^B - I_{z}{}^{l}\!_{k} P_l \psi^A ) +
\frac{\partial L}{\partial \beta}\, \beta \,\delta^{\mbox{\o}}_z = 0 \; . \label{eq:inv2}
\end{equation}
Here the latter term arises only for  the subgroup of dilatation (when $z =\mbox{\o} $). Absence of obvious dependence
of the Lagrangian density (\ref {eq:1017}) from $x^k $ yields the identity:
\begin{equation}
P_k L = \frac{\partial L}{\partial \psi^A} P_k \psi^A + \frac{\partial L}
{\partial  P_l  \psi^A}  P_k  P_l  \psi^A \; . \label{eq:inv1}
\end{equation}
Identities (\ref {eq:inv2}) and (\ref{eq:inv1}) are `strong' identities, which are satisfied independently of
validity of equations of the field $\psi^A$. When these equations are fulfilled, the given identities are equivalent
to existence of conservation laws. Indeed, it is possible to represent the variation of the action (\ref {eq:dJ0}) as
\begin{equation}
\delta J = \int_\Omega\, (dx) \left [\frac{\delta {\cal L}}{\delta \psi^A}
\,\bar{\delta} \psi^A + \frac{\partial {\cal L}}{\partial \beta}\,\bar{\delta}\beta +
\partial_\mu \left ({\cal L} \stackrel{\circ}{h}\!^{\mu}
\!_{k}\delta x^k - \stackrel{\circ}{h}\!^{\mu}\!_{k}\frac{\partial{\cal L}}
{\partial P_k \psi^A}\bar{\delta} \psi^A\right )\right ] = 0\;,\label{eq:varint}
\end{equation}
where the variational derivative has a standard structure. If the field equations, $\delta {\cal L}/\delta \psi^A = 0$,
are fulfilled, the variation (\ref {eq:varint}), with account of (\ref {eq:gM1}) and
$\partial_\mu \left (\sqrt{\mid \stackrel{\circ}{g}\mid}\stackrel{\circ}
{h}\!^{\mu}\!_{a} \right ) = 0$, is equal
\begin{equation}
\delta J = \int_\Omega\,(dx)\left ( \frac{\partial {\cal L}}{\partial \beta}\,\delta \beta +
\sqrt{\mid \stackrel{\circ}{g}\mid}\stackrel{\circ}{h}\!^{\mu}\!_k\,
\partial_\mu (a^l t^k\!_l + \omega^m M^k\!_{m} + \varepsilon \Delta^k )\right ) = 0\; , \label{eq:1203}
\end{equation}
where designations for an energy--momentum tensor $t^k\!_l$,  full $M^k\!_{m}$ and spin $S^k\!_{m}$
momemta, and also for full $\Delta^k $ and own $J^k$ dilatation currents of the field $\psi^A$ are introduced:
\begin{eqnarray}
&& t^k\!_l = L\delta^k_l - \frac{\partial L}{\partial P_k \psi^A}P_l \psi^A\;, \label{eq:ten}\\
&& M^k\!_{m} = S^k\!_{m} + I_{m}{}^{l}\!_{i}\,x^{i}\,t^k\!_{l}\; , \quad  S^k\!_{m} = -\,
\frac{\partial L}{\partial P_k \psi^A} I_{m}\!^{A}\!_{B} \psi^B \; , \label{eq:tsp}\\
&& \Delta^k = J^k - x^l\, t^k\!_l\; , \quad J^k = -\frac{\partial L}{\partial P_k \psi^A}w\psi^A\;.
\label{eq:tdl}
\end{eqnarray}
\par
Parameters $a^l $, $ \omega^m $ and $ \varepsilon $ are constant, but arbitrary, and the area of integration
$ \Omega $ is arbitrary. Therefore from identical equality to zero of a variation (\ref {eq:1203}) with the account
(\ref {eq:vg}), the following equalities follow in curvilinear system of coordinates:
\begin{eqnarray}
P_k\,t^k\!_l = 0 \;, \qquad P_k\,M^k\!_{m} = 0 \;, \qquad \sqrt{\mid \stackrel{\circ}{g}\mid}
\,P_k\,\Delta^k = \beta \frac{\partial {\cal L}} {\partial \beta} \;. \label{eq:dil}
\end{eqnarray}
\par
Equalities (\ref {eq:dil}) are the result of the first Noether theorem. The first two equalities yield the conservation
laws of the energy-momentum $t^k\!_l$ and the full momentum $M^k\!_{m}$ of the field $\psi^A$. For the conservation
of the dilatation current $\Delta^k$  it is necessary that an additional condition $\partial{\cal L}/\partial \beta = 0$
is fulfilled as consequence of the equation of the field $\psi^A$ (about the dilatational invariant Lagrangians with
explicit dependence on the parameter $\beta$ see \cite {Fr:book}, \cite {Fr-Sin}).
\par
Using the field equations, it is possible to show that the first equality (\ref {eq:dil}) is equivalent to the identity
(\ref {eq:inv1}), and the second and the third equalities (\ref {eq:dil}) are together equivalent to the identity
(\ref {eq:inv2}). Introducing designations $J^k\!_z= \{ S^k\!_m\,,\, J^k\,, 0\}$ and
\[
\stackrel{(\psi)}{\Theta}\!^k\!_z = J^k\!_z + X^l_z\,t^k\!_{l} = LX^k_z - \frac{\partial L}{\partial
P_k \psi^A} (I_z\!^{A}\!_{B} \psi^B + X_{z}^{l} P_l \psi^A ) \; ,
\]
it is possible to replace the equalities (\ref{eq:dil}) by one equality
\[
\partial_\mu \left (\sqrt{\mid \stackrel{\circ}{g}\mid}\stackrel{\circ}{h}\!^{\mu}\!_k\,\stackrel{(\psi)}
{\Theta}\!^k\!_z\right ) = \beta\,\frac{\partial {\cal L}}{\partial \beta}\,\delta^{\mbox{\o}}_z \;.
\]

\section {The principle of local invariance}
\setcounter {equation} {0}
\par
We suppose now the group ${\cal PW}(\omega\,,\,\varepsilon\,,\,a)$ as the localized group ${\cal PW}(x)$,
that is we consider its parameters $\{ \omega^{z} \} = \{ \omega^{m}\,, \, \varepsilon\,,\, a^k \}$
as arbitrary smooth enough (belonging to a class $C^2 $) functions of coordinates $\omega^z (x)$.
\par
Consider invariance of action integral (\ref {eq:1017}) under ${\cal PW}(x)$. Assuming the quantities $\omega^z(x)$ and
$\partial_\mu \omega^z(x)$ as arbitrary and independent functions of coordinates, from (\ref {eq:1203}) we obtain
conditions $t^{k}\!_{l} = 0\; ,\; M^{k}\!_{m}= 0\; ,\;\Delta^k = 0$. Thus action integral (\ref {eq:1017}) is
locally invariant then and only then, when the conservations laws are valid by virtue of the identical equality to zero
of the appropriate currents (\ref {eq:ten})--(\ref {eq:tdl}).
\par
It is possible to avoid this physically unsatisfactory result, if in the Lagrangian density (\ref {eq:1017}) enter some
additional fields named {\em gauged} (or {\em compensating}) having that property, that the
additional members, arising in action integral (\ref {eq:1017}) owing to transformation of a field $\psi^A$
under action of the localized group ${\cal PW}(x)$, will disappear by virtue of compensating them by
accordingly transformed gauge fields. Therefore gauge fields should be transformed under action $ {\cal PW} (x) $
as non-tensorial quantities extracting at this trans\-for\-ma\-ti\-on terms proportional a derivative from parameters
of the group ${\cal PW}(x)$. In case of the group ${\cal PW}(x)$, the new feature arises connected with the fact that
in this case the variation (\ref {eq:vg}) is equal
\begin{equation}
\delta g_{ij} = 2 \varepsilon(x) g_{ij}\;, \label{eq:ge}
\end{equation}
where $\varepsilon(x)$ is an arbitrary function. Therefore the metric òåíçîð becomes a function of a point of spacetime
and can be represented as
\begin{equation}
g_{ij} = \beta^{2}(x) g^{M}_{ij}\;, \label{eq:gM}
\end{equation}
that demands the account derivatives $P_k \beta(x)$ ($\beta(x)>0$) in the Lagrangian.
\par
The requirement of gauge invariance in application to the Poincar\'{e} group has been formulated in \cite {Fr1}, \cite {Fr2}
as some variational principle, which for a case of the localized Poincar\'{e}--Weyl group we generalize as follows.
\begin {quote}
{\bf Postulate 1} ({\em The principle of local invariance}). An action integral
\begin{equation}
J = \int_{\Omega}\,(d x)\,{\cal L}(\psi^A,\,P_k \psi^A,\,A^R_a,\,P_k A^R_a,\,
\beta(x)\,,\,P_k \beta(x) )\;, \label{eq:int}
\end{equation}
where the Lagrangian density ${\cal L}$ describes a field $\psi^A$, interaction of a field $\psi^A$ with an
additional gauge field $A^R_a$ and the free field $A^R_a$, is invariant under the action of the localized group
${\cal PW}(x)$, the gauge field being transformed as follows
\begin{equation}
\delta A^R_a = U^{R}_{za}\omega^z + S^{R\mu}_{za}\partial_{\mu}\omega^z \;, \label{eq:varA}
\end{equation}
where $U $ and $S $ are some matrix functions.
\end {quote}
\par
This variational principle allows to apply to gauge theories the first and the second Noether theorems and, in spite of
a generality of the formulation, it is sufficient to determine a structure of the Lagrangian density ${\cal L}$
and to find the matrix functions $U$, $S$. In the present work we generalize on the gauge theory of the
localized Poincar\'{e}--Weyl group the method of construction gauge theories, developed in \cite {Fr:book}--\cite {FrIv}
for the Poincar\'{e} group.
\par
The gauge fields equations, as well as the equations of the field $\psi^A$, are derived on the basis
of a principle of stationary action, which should be chosen as the second postulate.
\begin {quote}
{\bf Postulate 2} ({\em The principle of stationary action}). The equations of the field $\psi^A$ and the gauge
fields $A^R_a$ realize an extremum of the action integral (\ref {eq:int}), describing the field $\psi^A$, an interaction
of the field $\psi^A$ with the gauge field $A^R_a$ and the free gauge field $A^R_a$.
\end {quote}
\par From physical reasons it is necessary to conclude, that the full Lagrangian density ${\cal L}$ consists from
the Lagrangian density ${\cal L}_0$ of free gauge fields and from the Lagrangian density ${\cal L}_\psi $
describing the free field $\psi^A$ and the interaction of the field $\psi^A$ with gauge fields. Action integrals
for each of these Lagrangian densities separately need to be locally invariant, as it is natural to expect, that
the gauge field can exist irrespective of the field $\psi^A$.  We  formulate the given physical requirements
as the third postulate of the theory of gauge fields.
\begin {quote}
{\bf Postulate 3} ({\em An independent existence of a free gauge field}). The full Lagrangian density
${\cal L}$ of a physical system  depends additively from the locally invariant Lagrangian density ${\cal L}_0$
of free gauge fields: $ {\cal L} = {\cal L}_0 + {\cal L}_{\psi} $, where
\[
{\cal L}_0 = {\cal L}_0 (A^R_a\,,\,P_k A^R_a\,,\,\beta(x)\,,\,P_k \beta(x) )\;,\quad
\frac{\partial {\cal L}_0}{\partial \psi^A}= 0\;, \quad
\frac{\partial {\cal L}_0}{\partial P_k \psi^A} = 0\; .
\]
\end {quote}
\par
Further we shall always assume, that interaction of the field $\psi^A$ with gauge fields is carried out only through
connection with gauge fields, but not with their derivatives. In other words, all derivatives from the gauge fields
are contained only in the Lagrangian density ${\cal L}_0$. The interaction satisfying this condition is accepted to name
{\em minimal}. We  formulate this condition  as the forth postulate of the gauge fields theory.
\begin {quote}
{\bf Postulate 4} ({\em The principle of minimality of a gauge interaction}). In the Lagrangian density ${\cal L}_\psi$
of an interaction of a material field $\psi^A$ with gauge fields only derivatives from the material field $\psi^A$
are present. Thus the following conditions are satisfied,
\[
\frac{\partial {\cal L}_\psi}{\partial P_k A^R_a} = 0\;, \qquad
\frac{\partial {\cal L}_\psi}{\partial P_k \beta} = 0\; .
\]
\end {quote}
\par The variation of the action integral (\ref {eq:int}) with regard to the variation of the area $\Omega$ reads
\begin{eqnarray}
&& 0 = \delta J = \int_\Omega \,(dx)\, ((\partial_\mu \delta x^\mu )\,{\cal L} +
\delta {\cal L} ) = \int_{\Omega}\, (d x) \left (\frac{\delta {\cal L}}{\delta \psi^A}
\,\bar{\delta}\psi^A + \frac{\delta {\cal L}}{\delta A^R_a}\,\bar{\delta}
A^R_a + \frac{\delta {\cal L}}{\delta \beta}\,\bar{\delta}\beta\right )  + \nonumber \\
&& + \int_{\Omega}\, (dx)\, \partial_\mu \left ({\cal L}\stackrel{\circ}{h}\!^{\mu}\!_{k}
\,\delta x^k - \stackrel{\circ}{h}\!^{\mu}\!_{k} \frac{\partial {\cal L}}{\partial P_k \psi^A}
\,\bar{\delta} \psi^A - \stackrel{\circ}{h}\!^{\mu}\!_{k} \frac{\partial {\cal L}}
{\partial P_k A^R_a}\,\bar{\delta} A^R_a - \stackrel{\circ}{h}\!^{\mu}\!_{k} \frac{\partial
{\cal L}}{\partial P_k \beta}\,\bar{\delta}\beta \right ). \label{eq:varintA}
\end{eqnarray}
\par
As the action of the localized subgroup of dilatation $D_4$ yields (\ref {eq:ge}), the metric tensor becomes
a function and is subject to a variation. The principle of stationary action will be satisfied, if the variational
equations are valid,
\begin{equation}
\frac{\delta {\cal L}}{\delta \psi^A} = 0\; , \qquad \frac{\delta {\cal L}}{\delta A^R_a} = 0\;,
\qquad \frac{\delta {\cal L}}{\delta \beta(x)} = 0\; .
\label{eq:varurQA}
\end{equation}
\par
It is possible to show \cite {Fr:book} that the latter of these variational field equations is a consequence
of the others. We  assume that the equation of the field $\psi^A$ is always valid. According to the Postulate 3,
the full Lagrangian density of a gauge field consists from the Lagrangian density of a free gauge field and from the
Lagrangian density of interaction. For the separate interaction Lagrangian density the action integral is locally invariant,
but the variational equation of the gauge field $A^R_a$ is not satisfied. The equation of the field $A^R_a$ is valid
only for the full Lagrangian density ${\cal L}$. In this last case on the basis of (\ref {eq:varintA}), taking into
account (\ref {eq:varurQA}), (\ref {eq:varA}) and putting quantities
$\omega^z (x), \; \partial_\mu \omega^z (x), \; \partial_\mu\partial_\nu\omega^z (x)$
as arbitrary and independent functions of coordinates, we obtain a fundamental set of identities on extremals of the
fields $\psi^A $, $A^R_a $ and $ \beta (x) $:
\begin{eqnarray}
\partial_\mu (\stackrel{\circ}{h}\!^{\mu}\!_{k}\, \Theta^{k}\!_{z} ) = 0\;, \quad 
\stackrel{\circ}{h}\!^{\mu}\!_{k}\Theta^{k}\!_{z} - \partial_\nu {\cal M}^{\nu\mu}\!_{z} = 0\; ,
\quad {\cal M}^{(\nu\mu)}\!_{z} = 0\; , \label{eq:ft33}
\end{eqnarray}
where the following designations are introduced with regard to (\ref {eq:dbar}), (\ref {eq:1006a}) and (\ref {eq:varA}):
\begin{eqnarray}
&& \Theta^{k}\!_{z} = {\cal L}X^{k}_{z} - \frac{\partial {\cal L}}{\partial P_k \psi^A}
(I_{z}\!^{A}\!_{B}\psi^B + X^{l}_{z}P_l\psi^A ) -  \nonumber \\
&& - \frac{\partial {\cal L}}{\partial P_k A^R_a} (U^{R}_{za} + X^l_z P_l A^R_a ) -
\,\frac{\partial {\cal L}} {\partial P_k \beta}(\beta \delta^{\mbox{\o}}_z + X^l_z P_l \beta )\;, \nonumber\\
&& {\cal M}^{\nu\mu}\!_{z} =\, \stackrel{\circ}{h}\!^{\nu}\!_{k} \frac{\partial {\cal L}}
{\partial P_k A^R_a}\, S^{R\mu}_{za}\;. \label{eq:mA}
\end{eqnarray}
\par
The equalities (\ref {eq:ft33}) represent the relations of the second Noether theorem written down in curvilinear
system of coordinates. It is easy to be convinced that the first of these equality (representing the conservation law
of the appropriate current) is a consequence of two others. Thus it is shown, that introduction of gauge fields
leads to a dynamical realization of conservation laws. The quantity (\ref {eq:mA}) represents a superpotential for
the appropriate conservation current.

\section {Structure of the interaction Lagrangian }
\setcounter {equation} {0}
Following the method developed in \cite {Fr:book}, we  introduce the differential operator $M_R $:
\begin{equation}
M_R = \{ M_{m}\!^A\!_B\,,\, M_{\mbox{\o}}\!^A\!_B\,,\, M_k \!^A\!_B\}\; , \qquad R = \{m,\,\mbox{\o},\,k\}\;, \label{eq:MR}
\end{equation}
uniting the operators of full momemtum, full dilatation current and shift:
\begin{equation}
M_m\!^A\!_B = I_{m}\!^{A}\!_{B} + \delta^A_B \stackrel{\circ}{M}_m\; , \quad M_{\mbox{\o}}\!^A\!_B =
w\delta^A_B + \delta^A_B\stackrel{\circ}{D}\; , \quad M_k\!^A\!_B = \delta^A_B P_k\;.\label{eq:M}
\end{equation}
Also let us represent the gauge field $A^R_a $ as a set of three components:
\[
A^R_a = \{A^{m}_{a}\,,\, A_a\,,\,A^{k}_{a} \} \; ,
\]
where $A^m_a $ is the gauge field appropriate to the subgroup of 4-rotations ($r$-field), $A_a $ is the gauge
field of the subgroup of dilatation ($d$-field), and $A^k_a $ is the gauge field of the subgroup of translations
($t$-field) of the Poincar\'{e}--Weyl group.
\par
The following theorem on the structure of the Lagrangian density ${\cal L}_\psi $ of interaction of an external field
with the gauge fields, which represents a generalization on the Poincar\'{e}--Weyl group ${\cal PW}(x)$
the appropriate theorem proved in \cite {Fr:book} for a case of the Poincar\'{e} group.
\begin {quote}
{\bf Theorem 1.} There exists a gauge field $A^R_a $ with transformation structure (\ref {eq:varA}) of Postulate 1
under action of the localized Poincar\'{e}--Weyl group ${\cal PW}(x)$ and there are such matrix functions $Z $,
$U $ and $S $ of the gauge field, that the Lagrangian density
\begin{equation}
{\cal L_\psi} = \sqrt{\mid \bar g\mid}\, L_\psi(\psi^A,\,D_a \psi^A,\,\beta(x))\; , \quad
\sqrt{\mid \bar g\mid} = Z \sqrt{\mid \stackrel{\circ}{g}\mid}\;,\label{eq:LQ}
\end{equation}
satisfies to the principle of local invariance (Postulate 1) concerning the localized group ${\cal PW}(x)$,
${\cal L_\psi}$ being formed from the invariant concerning the non-localized group ${\cal PW}$ Lagrangian density
$L (\psi^A \, P_k \psi^A) $   by replacement of the differential operator $P_k $ on the gauge derivative operator
\begin{equation}
D_a = - A^{R}_{a} M_R \; ,\label{eq:AM}
\end{equation}
where the operator $M_R $ is given as (\ref {eq:MR}). Also the following representation of the gauge $t$-field
is valid:
\begin{equation}
A_a^k = D_a x^k \;.\label{eq:AD}
\end{equation}
\end {quote}
{\it Proof.} Substituting in (\ref {eq:AM}) the expression for the operator $M_R $, we  obtain
according to (\ref {eq:MR}) and (\ref {eq:M}) the explicit form of a gauge derivative for the group ${\cal PW}(x)$:
\begin{eqnarray}
&& D_a \psi^A = h^\mu\!_{a} \partial_\mu \psi^A - A^m_a I_{m}\!^{A}\!_{B} \psi^B -
w A_{a}\psi^A = h^{\mu}\!_{a}D_\mu \psi^A\;,  \label{eq:hD} \\
&& D_\mu \psi^A = \partial_\mu \psi^A - A^{m}\!_{\mu}I_{m}\!^{A}\!_{B} \psi^B - wA_\mu \psi^A\; . \label{eq:DmQ}
\end{eqnarray}
Here new quantities are introduced:
\begin{eqnarray}
&& h^{\mu}\!_{a} = \stackrel{\circ}{h}\!^{\mu}\!_{k} Y_a^k\; ,
\quad Y^k_a = A^R_a X^k_R = A^k_a + A^m_a I_{m}{}^{k}\!_{l}\,x^l - A_a x^k\; ,
\quad Z^a_k = (Y^{-1})^a_k \; , \label{eq:h} \\
&& h^a\!_{\mu} = (h^{-1})^a\!_{\mu} = Z^a_k \stackrel{\circ}{h}\!^{k}\!_{\mu}\;, 
\quad A^{m}\!_{\mu} =  A^{m}\!_{a}\,h^{a}\!_{\mu}\; , \quad A_\mu =  A_{a}\,h^{a}\!_{\mu}\;.\label{eq:YY}
\end{eqnarray}
\par By analogy to section 3, we  obtain the `strong' identities expressing conditions of invariance
of the action integral for the Lagrangian density (\ref {eq:LQ}) under action of the localized group ${\cal PW}(x)$.
The variation of action integral is equal
\begin{equation}
\delta \int_\Omega \,(dx)\,{\cal L}_{\psi} = \int_\Omega \,(dx)\Bigl (\sqrt{\mid \bar g\mid}\,
(\partial_\mu \delta x^\mu )\, L_{\psi} + \delta \left (\sqrt{\mid \bar g\mid}\right )\, L_{\psi} +
\sqrt{\mid \bar g\mid}\,\delta L_{\psi} \Bigr ) =0 \;. \label{eq:dJp}
\end{equation}
We  introduce a quantity $\bar g _{\mu\nu} $ (a tensor $g_{ab}$ has values of the tensor $g_{ij}$ (\ref{eq:gM1})):
\begin{equation}
\bar g_{\mu\nu} = g_{ab}h^a\!_{\mu}h^b\!_{\nu} = g_{ab}Z^a_k Z^b_l \stackrel{\circ}{h}\!^k\!_{\mu}
\stackrel{\circ}{h}\!^l\!_{\nu}\;,
\quad \bar g =\mbox{det}(\bar g_{\mu\nu}) = \stackrel{\circ}{g} Z^2\;, \quad Z = \mbox{det}(Z^a_k)\;, \label{eq:gmn}
\end{equation}
and also demand that matrixes $U $ and $S $ in the gauge fields transformation law (\ref {eq:varA}) were those
that for the quantity $g $ in (\ref {eq:gmn}) the following equality would be satisfied:
\begin{equation}
\delta \left (\sqrt{\mid \bar g\mid}\right ) = -\,\sqrt{\mid \bar g\mid}\,(\partial_\mu \delta x^\mu )\;.\label{eq:dsrg}
\end{equation}
The proof of the existence of the quantity $g_{\mu\nu} $ with the specified property is given at the end of the
section.
\par
Then the equality (\ref {eq:dJp}) by virtue of arbitrariness of the area $\Omega $ means $\delta L_{\psi} = 0 $:
\begin{eqnarray}
&& \delta L_{\psi} = \left (\frac{\partial L_\psi}{\partial \psi^{A}}\right )_{D\psi=const}
\delta\psi^A + \frac{\partial L_{\psi}}{\partial D_a \psi^A}\,
\delta D_a \psi^A + \frac{\partial L_{\psi}}{\partial \beta}\,\delta \beta = 0\; . \label{eq:vL}
\end{eqnarray}
Using (\ref {eq:hD}), (\ref {eq:h}), (\ref {eq:varA}), let us calculate a variation $\delta D_a \psi^A $
and then substitute it and also (\ref {eq:1022}) and (\ref {eq:ge}) into (\ref {eq:vL}). In the identity received
we collect factors before quantities $\omega^z(x)$ and $\partial_{\mu} \omega^z (x)$. In view of arbitrariness of
these quantities these factors separately should be equal to zero identically. The factor before
$\partial_{\mu} \omega^z (x) $ is equal
\[
\frac{\partial L_{\psi}}{\partial D_a \psi^A} (I_{R}\!^{A}\!_{B}\psi^B + X^{l}_{R}P_l\psi^A )
(S^{R\mu}_{za} - \delta^R_z h^{\mu}\!_{a}) = 0\;.
\]
This equality is satisfied identically at
\begin{equation}
S^{R\mu}_{za} = \delta^R_z h^{\mu}\!_{a}\;.\label{eq:S}
\end{equation}
Look over various sets of indexes, we find values of an unknown matrix $S $:
\begin{eqnarray}
&&S^{n\mu}_{ma} = \delta^n_m h^{\mu}\!_{a} \; , \quad  S^{n\mu}_{ka} = 0\; ,
\quad S^{k\mu}_{ma} = 0\; , \quad S^{l\mu}_{ka} = \delta^l_k h^{\mu}\!_{a}
\; , \label{eq:S1}\\
&&S^{\mu}_{a} = h^{\mu}\!_{a} \; , \quad  S_m{}^{\mu}_{a} = 0\; ,
\quad S^n{}^{\mu}_{a} = 0\; ,  \quad  S^k{}^\mu_a = 0\; ,
\quad S_k{}^\mu_a = 0 \; . \label{eq:S2}
\end{eqnarray}
\par
Now we take into account that an algebraic structure of the scalar $L_\psi $ should satisfy to the identity
(\ref {eq:inv2}), which owing to the replacement of $P_a$ by $D_a$ is equal
\[
\left (\frac{\partial L_\psi}{\partial \psi^{A}}\right )_{D\psi=const}
I_{z}\!^{A}\!_{B} \psi^B + \frac{\partial L_\psi}{\partial D_a \psi^A} (I_{z}\!^{A}\!_{B} D_a \psi^B -
I_{z}{}^{b}\!_{a} D_b \psi^A ) + \frac{\partial L_\psi}{\partial \beta}\,\beta\delta^{\mbox{\o}}_z = 0 \; .
\]
Considering the given identity, let us write out the factor at $\omega^z (x)$ in identity (\ref {eq:vL}). As a
result, we obtain some expression, which identically is equal to zero at the following set of matrixes $U $
in the transformation law (\ref {eq:varA}):
\begin{eqnarray}
&&U_{ma}^{n} = c_m\!^n\!_q A_{a}^{q} - I_m{}^b\!_a A_{b}^{n}\; ,
\quad  U_{ma} = - I_m{}^b\!_a A_b\;, \label{eq:Uz1}\\
&&U_{a}^{n} = A_{a}^{n}\; , \quad  U_{a} = A_a\;, \quad
U_{ka}^{n} = 0\; , \quad  U_{ka} = 0\; , \label{eq:Uz2}\\
&&U_{ma}^{k} = I_m{}^k\!_l A_{a}^{l} - I_m{}^b\!_a A_{b}^{k}\;, \quad
U_{ia}^{k}= -I_n{}^k\!_i A_{a}^{n} + \delta^k_i A_a \;, \quad
U^{k}_{a}= 0\;. \label{eq:Uz3}
\end{eqnarray}
These expressions can be expressed in short as
\begin{equation}
U_{za}^{R} = c_z\!^R\!_Q A_{a}^{Q} - I_z{}^b\!_a A_{b}^{R}\; , \label{eq:U}
\end{equation}
where each of the indexes $R$, $Q$, $z$ can take the values of the each indexes $m$, $k$, $\mbox{\o}$, and the
commutation relations (\ref {eq:CR}) of the Poincar\'{e}--Weyl group ${\cal PW}$ should be taken into account.
The expressions (\ref {eq:S}) and (\ref {eq:U}) found for an unknown function $Z $ and unknown matrix functions
$U $ and $S $, for which the Lagrangian density satisfies to identity (\ref {eq:dJp}), prove the basic statements
of the Theorem 1.
\par Now we shall prove the formula (\ref {eq:AD}). Owing to (\ref {eq:1004}), the quantity $x^k $ is transformed
as vector representation of the group ${\cal PW}$, and, comparing (\ref {eq:1004}) and (\ref {eq:1022}), we
find, that $w [x^k] =-1 $. At calculation a gauge derivative $D_a x^k $ we use the formulas (\ref {eq:hD}), (\ref {eq:h}),
(\ref {eq:1004}) and (\ref {eq:dx}):
\begin{eqnarray*}
&& D_a x^k = h^\mu\!_{a} \partial_\mu x^k - A^m_a I_{m}{}^{k}{}_{l} x^l - w[x^k] A_{a}x^k = \nonumber\\
&& = \stackrel{\circ}{h}{}^{\mu}{}_{i} (A^i_a + A^m_a I_{m}{}^{i}{}_{l}\,x^l - A_a x^i )\partial_\mu x^k -
 A^m_a I_{m}{}^{k}{}_{l} x^l + A_{a}x^k = A^k_a\;.
\end{eqnarray*}
The formula (\ref {eq:AD}) clears up a geometrical sense of the gauge field of the subgroup of translations.
This formula generalizes on the Poincar\'{e}--Weyl group the similar formula, which arises at the gauge approach for
the Poincar\'{e} group \cite {Fr:book}--\cite {FrIv}.
\par
 Let us find transformation laws of components of the gauge field under action of the localized Poincar\'{e}--Weyl
 group ${\cal PW}(\omega\,,\,\varepsilon\,,\,a)$. The general form of the transformation law is
 determined on account of the principle of local invariance by the expression (\ref {eq:varA}). We break in this
 expression indexes $R $ and $z $ on three indexes concerning subgroups of 4-rotations, dilatation and
 translations and substitute in these formulas the concrete values of the matrix functions $U $
 (\ref {eq:Uz1})--(\ref {eq:Uz3}) and $S $ (\ref {eq:S1}), (\ref {eq:S2}). As a result we obtain the general rules of
 transformations for the fields $A^m_a $, $A_a $ and $A^k_a $:
 \begin{eqnarray}
&&\delta A^m_a = \omega^{n} c_{n}\!^m\!_q A_{a}^{q} - \omega^{n}I_n{}^b\!_a A_{b}^{m}
+ \varepsilon A^m_a + h^{\mu}\!_{a}\,\partial_{\mu}\omega^{m}\;, \label{eq:delAm}\\
&&\delta A_{a} = -\omega^{n}I_n{}^l\!_a A_l + \varepsilon A_a
+ h^{\mu}\!_{a}\,\partial_{\mu}\varepsilon \; ,\label{eq:delAa}\\
&&\delta A_{a}^{k} = \omega^{n} (I_n{}^k\!_l A_{a}^{l} - I_n{}^l\!_a A_{l}^{k})
+ a^l (- A^n_a I_n{}^k\!_l + A_a \delta^k_l)  + h^{\mu}\!_{a}\,\partial_{\mu} a^{k}
\; .  \label{eq:delAk}
\end{eqnarray}
\par
A variation of the quantity $h^{\mu}\!_{a}$ is determined on the basis of formulas (\ref {eq:h}), the transformation
law\footnote [1]{The transformation law (\ref {eq:dY}) is externally similar to the transformation law of gauge fields
(\ref {eq:varA}). However, parameters of this transformation $\delta x^{k}$ do not bear in themselves (without
their concrete definition) any information on the group ${\cal PW}(x)$. Therefore the quantity $Y_{a}^{k}$ is
impossible to consider as a gauge field for the group ${\cal PW}(x)$. The quantity $Y_{a}^{k}$ is closely connected
with the tetrads $h^\mu\!_a$.} of the quantity $Y_{a}^{k}$ having been determined previously on the basis of the
variations (\ref {eq:delAm})--(\ref {eq:delAk}):
\begin{equation}
\delta Y_{a}^{k} = - \omega^n I_n{}^l\!_a Y_{l}^{k} + \varepsilon Y_{a}^{k} +
h^{\mu}\!_{a} \,\partial_\mu\delta x^{k} \; .\label{eq:dY}
\end{equation}
As a result one finds
\begin{equation}
\delta h^\mu\!_a = - \omega^{n}I_n{}^b\!_a\, h^\mu\!_b + \varepsilon h^\mu\!_a
+ h^\nu\!_a \,\partial_\nu \delta x^\mu\; . \label{eq:dhma} \\
\end{equation}
At last, from (\ref {eq:YY}) we find transformation rules of the quantities $h^a\!_\mu $, $A^m\!_\mu $ and $A_m$:
\begin{eqnarray}
&&\delta h^{a}\!_{\mu} = \omega^n I_n{}^a\!_b\, h^{b}\!_{\mu} - \varepsilon h^{a}\!_{\mu}
- h^a\!_\nu \,\partial_\mu \delta x^\nu \; ,  \label{eq:dham} \\
&&\delta A^m\!_\mu = \omega^{n} c_{n}\!^m\!_q A_{\mu}^{q} + \partial_\mu \omega^m
- A^m\!_\nu \,\partial_\mu\delta x^\nu \; ,  \nonumber\\
&&\delta A_\mu = \partial_\mu \varepsilon - A_\nu \,\partial_\mu\delta x^\nu \;.\nonumber
\end{eqnarray}
\par
After the geometrical interpretation of the theory (section 7) the quantity $h^\mu\!_a$ will be interpreted as a tetrad
potential. We see, that at localization of the group ${\cal PW}$ (as well as in case of the Poincar\'{e} group
\cite {Fr:book}), the quantity $h^\mu\!_a$ is not a gauge field because it is transformed as tensor
(in agree with (\ref {eq:dhma}) and (\ref {eq:dham})), against the gauge fields $A^m_a $, $A_a $ and
$A^k_a $, which are transformed concerning the group ${\cal PW}(x)$ by non-tensorial rules. We note, that in
\cite {Iv-Sar3}, \cite {Sard} it was specified the fact, that tetrads are not true potentials of a gravitational
field.
\par We shall prove now validity of the formula (\ref {eq:dsrg}) for components of the quantity $g_{\mu\nu}$
determined by (\ref {eq:gmn}). Determined by formulas (\ref {eq:S1})--(\ref {eq:Uz3}), the expressions for matrix
functions $S $ and $U $ yield the transformation law (\ref {eq:dham}) for the quantity $h^a\!_\mu$. Owing to
the definition (\ref {eq:gmn}), we have
\[
\delta \left (\sqrt{\mid \bar g\mid}\right ) = \frac{1}{2\sqrt{\mid \bar g\mid}}\,\delta \mid \bar g\mid\, =
\frac{1}{2\sqrt{\mid \bar g\mid}}\mid \bar g\mid\,g^{\mu\nu}\delta g_{\mu\nu} =
\frac12 \sqrt{\mid \bar g\mid}\, g^{\mu\nu} (h^a\!_{\mu}h^b\!_{\nu}\delta g_{ab} + 2g_{ab}h^b\!_{\nu}\delta h^a\!_{\mu})\;.
\]
Substituting here the expressions for variations of the quantities $g _ {ab} $ and $h^a \! _ {\mu} $ from (\ref {eq:vg})
and (\ref {eq:dham}), we find
\begin{eqnarray*}
&&\delta \left (\sqrt{\mid \bar g\mid}\right ) = \,\frac12 \sqrt{\mid \bar g\mid}\, g^{\mu\nu}
\Bigl (2\varepsilon (x)g_{ab} h^a\!_{\mu}h^b\!_{\nu} + 2g_{ab}h^b\!_{\nu}\,(\omega^n (x)I_n{}^a\!_c\, h^{c}\!_{\mu} -
\varepsilon (x)h^{a}\!_{\mu} - \nonumber\\
&& - h^a\!_\sigma \,\partial_\mu \delta x^\sigma) \Bigr ) = \frac12 \sqrt{\mid \bar g\mid}\,(8\varepsilon (x) +
\omega^n (x)I_n{}^a\!_a - 8\varepsilon (x) - 2\partial_\mu \delta x^\mu ) =
- \sqrt{\mid \bar g\mid}\,\partial_\mu \delta x^\mu \; .
\end{eqnarray*}
Thus it is proved, that such matrix functions $S $ and $U $ exist, namely, given by the expressions
(\ref {eq:S1})--(\ref {eq:Uz3}), for which the equality (\ref {eq:dsrg}) is satisfied. $ \bigsqcup $

\section {Free gauge field Lagrangian. Equations of gauge fields}
\setcounter {equation} {0}
For definition of structure of the free gauge field Lagrangian for the group ${\cal PW}(x)$ we need to find such
functions of the gauge fields, which are transformed concerning the group ${\cal PW}(x)$ as tensors. Knowing, that
the gauge derivative $D_a \psi^A $ is tensor, we calculate the commutator of gauge derivatives of a field $\psi^{A}$:
\[
2D_{[a}D_{b]}\psi^{A} = - F^{m}\!_{ab}I_{m}\!^{A}\!_{B}\psi^{B}
+ w F_{ab}\psi^{A} - F^c\!_{ab}D_c \psi^{A} \; .
\]
The quantities $F^m\!_ {ab}$, $F_{ab}$ and $F^c\!_{ab}$ are tensors. They are defined by expressions
\begin{eqnarray}
&& F^{m}\!_{ab} = 2h^\lambda\!_{ [a} \partial_{\mid\lambda\mid} A^m_{b]} +
 A^m_c C^c\!_{ab} - c_n\!^m\!_q A^n_a A^q_b\; ,\label{eq:Fab} \\
&& F_{ab} = 2h^\lambda\!_{ [a} \partial_{\mid\lambda\mid} A_{b]} +
 A_c C^c\!_{ab}\; ,\label{eq:F} \\
&&F^{c}\!_{ab} = C^c\!_{ab} + 2I_n\!^c\!_{ [a} A^n_{b]} + 2 A_{[a}\delta^{c}_{b]}\;,
\label{eq:Tab}\\
&& C^c\!_{ab} = - 2h^c\!_\tau h^\lambda\!_{ [a}
\partial_{\mid\lambda\mid} h^\tau\!_{b]} = 2 h^\lambda\!_a h^\tau\!_b
\partial_{ [\lambda} h^c\!_{\tau ]}\;. \nonumber
\end{eqnarray}
Besides, the following contraction of a gauge derivative is also a vector :
\begin{equation}
Q_{a} = - g^{bc}D_a g_{bc} = - h^\mu\!_a\, g^{bc}\partial_{\mu} g_{bc} - 2g^{bc}A^z_a I_{zbc} =
8(A_a - h^\mu\!_a\,\partial_{\mu}\ln\beta(x))\;. \label{eq:Qa}
\end{equation}
With the help of the variations of the gauge fields (\ref {eq:delAm})--(\ref {eq:delAk}) and also of the variation of
the tetrads (\ref {eq:dhma}), (\ref {eq:dham}), it is possible to show by explicit calculations that
(\ref {eq:Fab})--(\ref {eq:Tab}), (\ref {eq:Qa}) are transformed as covariant quantities under action of the group
${\cal PW}(x)$:
\begin{eqnarray*}
&&\delta F^m\!_{ab} = \omega^{n} (c_{n}\!^{m}\!_{q} F^q\!_{ab} - I_n{}^c\!_a F^m\!_{cb} -
I_n{}^c\!_b F^m\!_{ac} )  + 2\varepsilon F^m\!_{ab}\; ,\\
&&\delta F_{ab} = -\omega^{n} (I_n{}^c\!_a F_{cb} + I_n{}^c\!_b F_{ac} )+ 2\varepsilon F_{ab}\; ,\\
&&\delta F^c\!_{ab} = \omega^{n} (I_n{}^c\!_d F^d\!_{ab} - I_n{}^d\!_a F^c\!_{db} -
I_n{}^d\!_b F^c\!_{ad} ) + \varepsilon F^c\!_{ab}\; ,\\
&&\delta Q_{a} = -\,\omega^{n} I_n{}^b\!_a Q_{b} + \varepsilon Q_{a} \; .
\end{eqnarray*}
These tensor quantities contain derivatives from gauge fields only of the first degree, therefore it is natural to name them
as gauge field strengthes. For construction of the free gauge field Lagrangian density it is necessary to use the scalars
formed from the gauge field strengthes. As a result we come to a conclusion about the structure of the free gauge field Lagrangian
density.
\begin{quote}
{\bf Theorem 2.} The following Lagrangian density
\begin{equation}
{\cal L}_{0} = \sqrt{\mid \bar g\mid}\,L_{0}(F^{m}\!_{ab},\,F_{ab},\,F^{c}\!_{ab},\,Q_{a},\,\beta(x))\; ,
\label{eq:L0FT} \end{equation}
where $L_{0}$ is a scalar function formed from the gauge field strengthes  (\ref {eq:Fab})--(\ref {eq:Tab}),
(\ref {eq:Qa}), satisfies to the principle of local invariance.
\end{quote}
\par
A full Lagrangian density of a set of a field $\psi^{A}$ and a gauge field is
\begin{equation}
{\cal L} = {\cal L}_{0} + {\cal L}_{\psi} \; , \label{eq:Ltot}
\end{equation}
where ${\cal L}_{0}$ is given by the expression (\ref {eq:L0FT}), and $ {\cal L} _ {\psi} $ -- by the
expression (\ref {eq:LQ}). For the full Lagrangian density (\ref {eq:Ltot}) the variational gauge field equations
(\ref {eq:varurQA}) are satisfied:
\begin{equation}
\frac{\delta{\cal L}_0}{\delta A_{a}^{m}} = - \frac{\partial {\cal L}_{\psi}}
{\partial A_{a}^{m}}\, , \quad \frac{\delta{\cal L}_0}{\delta A_{a}} =
- \frac{\partial {\cal L}_{\psi}}{\partial A_{a}}\, , \quad \frac{\delta{\cal L}_0}
{\delta A_{a}^{k}} = - \frac{\partial {\cal L}_{\psi}}{\partial A_{a}^{k}}\, , \quad
\frac{\delta{\cal L}_0}{\delta \beta(x)} = -\frac{\delta{\cal L}_{\psi}}{\delta \beta(x)}\;.
\label{eq:urAm}  \end{equation}
As it has been already pointed out, the latter of these variational field equations is a consequence of  the others.
The right parts of these equations represent sources of the gauge fields, which are calculated with the help
of expression (\ref {eq:LQ}):
\begin{eqnarray}
&&-\frac{\partial {\cal L}_{\psi}} {\partial A^m_{a}} = \sqrt{\mid \bar g\mid}\,
(\stackrel{\circ}{M}\!^a\!_{m} + S^a\!_{m} )\; , \nonumber \\
&&\stackrel{\circ}{M}\!^a\!_{m} = I_m{}^c\!_b\, x^b \stackrel{(g)}{t}\!^a\!_c\;, \quad
\sqrt{\mid \bar g\mid}\, S^a\!_{m} = \frac{\partial{\cal L}_{\psi}}
{\partial D_a \psi^{A}}I_{m}\!^{A}\!_{B}\psi^{B}\; ,\label{eq:istAm} \\
&& - \frac{\partial {\cal L}_{\psi}} {\partial A_{a}} = \sqrt{\mid \bar g\mid}\,\Delta^{a}\;, \;\;
\Delta^{a} = -x^{b}\stackrel{(g)}{t}\!^a\!_{b} + J^a\; , \;\;  \sqrt{\mid \bar g\mid}\, J^a =
\frac{\partial{\cal L}_{\psi}}{\partial D_a \psi^{A}} w \psi^{A}\, , \label{eq:istA} \\
&& - \frac{\partial {\cal L}_{\psi}} {\partial A_{a}^{k}} = \sqrt{\mid \bar g\mid}\, \stackrel{(g)}{t}\!^a\!_k\; ,
\quad \sqrt{\mid \bar g\mid} \stackrel{(g)}{t}\!^{a}\!_k = Z^a_l \,\left ({\cal L}_{\psi}\delta^l_k -
\frac{\partial{\cal L}_{\psi}}{P_l \psi^A }\, P_k \psi^A \right )\; . \label{eq:istAk}
\end{eqnarray}
Here $\stackrel {(g)}{t}\!^{a}\!_k$ is an energy-momentum tensor, $S^a\!_{m}$ is a spin,
$\stackrel{\circ}{M}\!^a\!_{m}$ is an orbital momenta of an external field $\psi^{A}$ \cite {Bog-Sch},
$\Delta^{a}$ is full, and $J^a$ is a proper dilatation currents of $\psi^{A}$ \cite{MacSal}.
\par
The equations (\ref {eq:urAm})--(\ref {eq:istAk}) allow to generalize on the localized group ${\cal PW}(x)$ the
theorem on the sources of gauge fields, proved in \cite {Fr1}, \cite {Fr:book} for the localized Poincar\'{e} group.
\begin {quote}
{\bf Theorem 3} ({\em on the sources of gauge fields}). Sources of the gauge fields introduced by the localized
Poincar\'{e}--Weyl group ${\cal PW}(x)$, are invariants of the Noether theorem appropriate to the non-localized
Poincar\'{e}--Weyl group ${\cal PW}$.
\end {quote}
According to this theorem, the dilatation current $\Delta^{a}$ of an external field (connected with possible
presence of a dilatation charge in nature \cite {BabDil}--\cite {CQG-03}) arises as a source of the gauge field in
addition to the currents introduced by the localized  Poincar\'{e} group.
\par
As an example let us consider a case, when an external field is a spinor field. Then, according to the
Theorem 1, a spinor field Lagrangian interacting with the gauge fields of the Poincar\'{e}--Weyl group reads
\[
{\cal L_\psi} = \sqrt{\mid \bar g\mid}\, L_\psi\; , \quad L_{\psi} =
\bar{\psi}_A \gamma^a D_a \psi^A - m\,\bar{\psi}_A \psi^A \;,
\]
where the gauge derivative $D_a $ is determined by expressions (\ref {eq:hD}) and (\ref {eq:h}). We consider
the gauge fields as weak fields and pass to a Cartesian system of coordinates, for which
$\stackrel{\circ}{h}{}^{\mu}{}_{a} = \delta^\mu_a $. Substituting the expressions for $D_a $ in the Lagrangian,
we receive
\begin{eqnarray}
&& L_{\psi} = \bar{\psi}_A \gamma^a D^{(t)}_a \psi^A - m\,\bar{\psi}_A \psi^A + L_{\psi M} + L_{\psi D}\;,\label{eq:Lps1} \\
&& D^{(t)}_a \psi^A = A^k_a \partial_k \psi^A - A^m_a I_{m}\!^{A}\!_{B} \psi^B - wA_{a}\psi^A \;,  \label{eq:Dt} \\
&& L_{\psi M} = A^m_a I_{m}\!^{k}\!_{l}\, x^l \bar{\psi}_A \gamma^a \partial_k \psi^A\;, \qquad
L_{\psi D} = -A_a x^k \bar{\psi}_A \gamma^a \partial_k \psi^A\;. \label{eq:LMD}
\end{eqnarray}
Here the first term contains a derivative $D^{(t)}_a \psi^A $, which appears from (\ref {eq:hD}) by exchange
the tetrads $h^\mu\!_a$ for the translation gauge field $A^k_a$. This term contains a direct interaction of a
gravitational field with the spin momentum of $\psi^A $, which is standard for General Relativity. The second term
describes a direct interaction of the Lorentz gauge field with the orbital momentum of the spinor field. The third
term describes a direct interaction of the dilatation gauge field with the orbital dilatation current. Both these
interactions are absent in all those theories, in which the tetrads do not represent as (\ref {eq:h}), in particular,
in the theories advanced in \cite {Char-Ta}, \cite {Kas} and also in General Relativity.
\par
Let us calculate the second term for example. To that end we calculate the expression (\ref {eq:istAk}) for the
energy-momentum tensor of the spinor field:
\[
t^a \! _ k = Z^a_k \, L _ {\psi} - \bar {\psi} _A \gamma^a \partial_k \psi^A \;,
\]
and also use the expression for generators of vector representation of the group ${\cal PW}(x)$:
\begin{equation}
I_{ij}\!^{a}\!_{b} = \delta^{a}_{i}g_{jb} - \delta^{a}_{j}g_{ib}\quad (m\rightarrow \{i,j\},\; i < j)\; .\label{eq:Ikl}
\end{equation}
As a result we obtain:
\[
L_{\psi M} = (1/2)A^{ij}_a M^a_{ij}\;, \qquad M^a_{ij} = x_i\,t^a\!_j - x_j\,t^a\!_i\;,
\]
where it is taken into account that $L _ {\psi} =0 $ by virtue of the spinor field equation.
\par
The given interaction describes the effect, which will become apparent, in particular, in a precession of electronic
orbits under action of an external gravitational field and theoretically can be used for detecting gravitational waves.

\section {Interactions of gauge fields}
\setcounter {equation} {0}

Let's consider a full group of gauge symmetries $\Gamma (x)$, into which the group ${\cal PW}(x)$ enters as
a component of a direct product. As an example, we shall consider a group
\[
\Gamma (x)= {\cal PW}(x)\bigotimes SU_3(x)\bigotimes U_1(x)\;,
\]
where $SU_3(x)$ is the non-Abelian gauge color group of quantum chromodynamics, and $U_1 (x) $ is the gauge group of
electrodynamics. Then, applying to the group $\Gamma (x)$ the general theory of gauge fields \cite {Fr:book}, we
obtain, according to the Theorem 2, that a strength tensor of a unified gauge field reads
\begin{equation}
F^{M}\!_{ab} = 2h^\lambda\!_{ [a} \partial_{\mid\lambda\mid} A^M_{b]} +
 A^M_c C^c\!_{ab} - c_N\!^M\!_Q A^N_a A^Q_b\; ,\label{eq:FM}
\end{equation}
where indexes $M, \, N, \, Q $ run the values of indexes of all infinitesimal operators of all components of the direct
product. Further it is necessary to take into account, that the various infinitesimal operators of components of the direct
product commute. As a result, structural constants and a metric òåíçîð $g_{MN}$ of a group space, and also squares
of the tensor (\ref {eq:FM}) break up to the blocks concerning everyone of a component of the direct product.
\par
Then the gauge field strength tensor (\ref {eq:FM}) will be represented  as a set of components
$F^{M}\!_{ab} = \{F^{m}\!_{ab},\,F^{i}\!_{ab},\,F_{ab}\}$. Here a tensor $F^{m}\!_{ab}$ concerns to the group
${\cal PW}(x)$ and is given by the expression (\ref {eq:Fab}). A tensor $F^{i}\!_{ab}$ describes a color gauge field,
and a tensor $F_{ab}$ describes an electromagnetic field. These tensors are given by expressions
\begin{equation}
F^{i}\!_{ab} = 2h^\lambda\!_{ [a} \partial_{\mid\lambda\mid} A^i_{b]} +  A^i_c C^c\!_{ab} - c_k{}^i{}_l A^k_a A^l_b\; ,
\qquad F_{ab} = 2h^\lambda\!_{ [a} \partial_{\mid\lambda\mid} A_{b]} +  A_c C^c\!_{ab}\; ,\label{eq:FÑÅ}
\end{equation}
where $c_k{}^i{}_l $  are structural constants of the group $SU_3 $. Formulas (\ref {eq:FÑÅ}) describe interaction of
color and electromagnetic fields with the gauge field of the group ${\cal PW}(x)$.
\par
In connection with stated, it is erroneous to describe the given interactions by the replacement of usual derivatives on
gauge covariant ones:
\begin{equation}
F^{i}\!_{ab} = 2D_{[a} A^i_{b]} +  A^i_c C^c\!_{ab} - c_k{}^i{}_l A^k_a A^l_b\; ,
\qquad F_{ab} = 2D_{[a} A_{b]} +  A_c C^c\!_{ab}\; ,\label{eq:MÑÅ}
\end{equation}
that it is often accepted to use.

\section {Geometrical interpretation}
\setcounter {equation} {0}
Well-known, that the theory of gauge fields can be interpreted in terms of differential geometry and of fiber
bundles (see \cite {Sl-Fad}, \cite {Iv-Sar3}--\cite {Sard}, \cite {Lub}, \cite {Kon-Pop} and the literature cited
there). Already in the paper of Weyl \cite {Weyl} a generalization of Riemann geometry to Weyl geometry was found
as a consequence of the requirement of invariance of the theory concerning local change of scales. A fiber bundles
treatment of Weyl geometry one can found in \cite{Foll}.
\par
In \cite {Kib1}--\cite {He:rmf}, \cite {Fr:book} it was shown, how as a consequence of localization of the Poincar\'{e}
group the Riemann--Cartan geometry arises. Let us show that the results of the previous sections can be interpreted as
a realization the Weyl--Cartan differential geometry on the spacetime manyfold $\cal M$. In a basis of the given
geometrical interpretation there contains
the identification of the gauge derivative (\ref {eq:DmQ}) and a covariant derivative on the spacetime manyfold,
and also the interpretation of a frame $\vec e_ {a}$ as an orthogonal frame of the space (tangent to the manyfold
$\cal M$), in which the localized Poincar\'{e}--Weyl group operates:
\begin{equation}
\vec e_{a} = \vec e_{\mu} h^{\mu}\!_a \; , \qquad \{\vec e_{\mu}\} = \{\partial_\mu\}\;,
\qquad \left [\vec e_{a},\;\vec e_{b}\right ] = - C^{c}\!_{ab}\vec e_{c}\; , \quad
C^c\!_{ab} = 2 h^\lambda\!_a h^\tau\!_b\partial_{ [\lambda} h^c\!_{\tau ]}\;.
\label{eq:2203}
\end{equation}
Thus the basis $\vec e_{a} $ is a non-holonomic basis, and a quantity $C^{c}\!_{ab}$ is an object of
non\-ho\-lo\-no\-mi\-ty \cite {Sh-Str}. The shift operator $P_{a}$ is redefined: $P_{a} = h^{\mu}\!_{a}\partial_{\mu}$,
$\left [P_{a},\; P_{b} \right] = - C^{c}\!_{ab} P_{c} $, and represents the shift operator of a tangent space.
The metric tensor $g _ {ab} $ appears to be a metric tensor of a tangent space, which element of length is equal
$dx^a = h^a\!_{\mu} dx^\mu $. Then the square of an element of length of this space will be equal
\[
ds^2 = g_{ab}dx^a dx^b = g_{ab}h^a\!_{\mu}h^b\!_{\nu}dx^\mu dx^\nu = \bar g_{\mu\nu}dx^\mu dx^\nu \; ,
\]
that allows to interpret quantities $\bar g_{\mu\nu} $, calculated under the formula (\ref {eq:gmn}), as
components of a Weyl--Cartan metric tensor of spacetime manifold $\cal {M}$ in coordinate holonomic basis:
\begin{eqnarray}
&&\bar g_{\mu\nu} = \breve g(\vec e_{\mu},\,\vec e_{\nu}) = \breve g(\vec e_{a},\,\vec e_{b})\,h^{a}\!_{\mu}h^{b}\!_{\nu}
 = g_{ab}h^{a}\!_{\mu}h^{b}\!_{\nu} = \beta^2\,g_{\mu\nu}\;, \quad g_{\mu\nu} = g^{M}_{ab}h^{a}\!_{\mu}h^{b}\!_{\nu} \;.
\label{eq:gWC}\\
&& \sqrt{\mid \bar g\mid} = \beta^4\,\sqrt{\mid g\mid} = \beta^4\,h\;, \quad g = \det{(g_{\mu\nu})}\;, \quad
h = \det{(h^a\!_\mu)}\;. \label{eq:hdet}
\end{eqnarray}
At such interpretation of the quantities $\bar g_{\mu\nu}$, the formula (\ref {eq:dsrg}) becomes obvious. The quantities
$g_{\mu\nu} $ are the coordinate holonomic components of a Riemann--Cartan metric tensor of spacetime manifold $\cal {M}$.
Two types of indexes, arising in the theory -- tetrad $a,\;b,\;\ldots $ and coordinate $\mu,\;\nu,\;\ldots $, change
each other by means of the quantities $h^a\!_\mu $, which are interpreted as tetrads. It is generally conventional that
contractions of quantities with Greek holonomic indexes are performed with the Riemann--Cartan metric tensor $g_{\mu\nu}$.
\par
In the formula for a gauge derivative (\ref {eq:DmQ}) it is necessary to use the expression (\ref {eq:Ikl}) for
generators $I_m\!^A\!_B$ of a vector representation $\psi^{A} =v^{a}$ of the Poincar\'{e}--Weyl group, a weight
of a vector field being equal $w [v^{a}] =-1 $. Then, equating the expression for a gauge derivative from a vector:
\[
D_\mu v^a = \partial_\mu v^a  - A^{m}\!_{\mu}I_{m}\!^{a}\!_{b} v^b + A_\mu v^a \; ,
\]
to the expression for the covariant derivative of a vector in differential geometry:
$\nabla_\mu v^a = \partial_\mu v^a + \Gamma^a\!_{b\mu } v^b$, we find connection coefficients in non-holonomic basis:
\begin{equation}
\Gamma^{a}\!_{b\mu} = - A^{m}\!_{\mu}I_{m}\!^{a}\!_{b} + \delta^a_b A_\mu \; .\label{eq:Gab}
\end{equation}
\par
With the purpose to define a covariant derivative $\nabla_\mu$ for quantities with coordinate indexes, it is postulated, that
\[
\nabla_{\lambda} h^a\!_\mu = \partial_{\lambda}h^a\!_\mu + \Gamma^a\!_{b\lambda} h^b\!_\mu
- \Gamma^\nu\!_{\mu\lambda} h^a\!_\nu = 0\;.
\]
From this formula we find connection coefficients in holonomic coordinate basis:
\begin{equation}
\Gamma^\lambda\!_{\nu\mu} = h^\lambda\!_a h^{b}\!_{\nu}\Gamma^{a}\!_{b\mu}
+ h^\lambda\!_{a}\partial_{\mu} h^{a}\!_{\nu}\; . \label{eq:2217}
\end{equation}
\par According to (\ref {eq:ge}), under action of the group ${\cal PW}(x)$ the metric tensor of a tangent space is
multiplied by an arbitrary function and can be represented as (\ref {eq:gM}) \cite {Ut2}--\cite {Ut3}. Calculating
a variation of this expression (\ref {eq:gM}) and comparing to a variation of the metric tensor (\ref {eq:ge}), we find
that $ \delta\beta =\beta\varepsilon (x) $. Thus the field $\beta (x)$ has the weight $w[\beta (x)] =1 $. This field
coincides with the scalar field introduced by Dirac \cite {Dir}, and can be represented as $\beta (x) = \exp{\sigma (x)}$,
where $\sigma (x)$ is a {\em dilaton} field. The field $\beta (x)$ is also similar to the `measure' scalar field
introduced by Utiyama \cite {Uti}. In essence the field $\beta (x)$ is a factor of components of the tangent space
metric tensor in Weyl--Cartan geometry.
\par
It is known from differential geometry \cite {Sh-Str} that a nonmetricity tensor is equal
\[
Q_{ab\mu} = - \nabla_{\mu}g_{ab} = -\partial_{\mu}g_{ab} + 2\Gamma_{(ab)\mu}\;.
\]
Substituting here and also in (\ref {eq:Qa}) the expression (\ref {eq:gM}), we obtain
\begin{equation}
Q_{ab\mu} = \frac{1}{4}g_{ab}Q_\mu\;, \quad Q_{\mu} = g^{ab}Q_{ab\mu}\;, \quad Q_{\mu} = 8 (A_{\mu} -
\partial_{\mu}\ln \beta(x)) = Q_a h^a\!_\mu \;.
\label{eq:nem} \end{equation}
If the nonmetricity tensor satisfies to first two equalities (\ref {eq:nem}) then nonmetricity is Weyl's nonmetricity.
In this case a trace $Q _{\mu}$ of the nonmetricity tensor is named Weyl vector, it is expressed through the vector
(\ref {eq:Qa}).
\par
Let us introduce quantities
\begin{eqnarray}
&& F^m\!_{\mu\nu} = F^m\!_{ab}h^a\!_\mu h^b\!_\nu = 2\partial_{ [\mu}
A^m\!_{\nu ]} - c_n\!^m\!_q A^n\!_\mu A^q_\nu \; ,\label{eq:Fmmn} \\
&&F_{\mu\nu} = F_{ab}h^a\!_\mu h^b\!_\nu = 2\partial_{[\mu}A_{\nu ]}\;,\label{eq:Fmunu} \\
&& F^c\!_{\mu\nu} = F^c\!_{ab} h^a\!_\mu h^b\!_\nu = 2\partial_{ [\mu} h^c\!_{\nu ]} +
2I_n{}^c\!_a h^a\!_{ [\mu} A^n_{\nu]} + 2A_{[\mu}h^{c}_{\nu ]}\; ,\label{eq:Tmunu}
\end{eqnarray}
which  represent (together with (\ref {eq:nem})) the gauge field strengthes for a new set of dynamic variables
$\{A^m\!_\mu,\,A_\mu,\,h^a\!_\mu,\,\beta(x)\}$.
\par
Substituting in the curvature of spacetime manyfold the expression for connection coefficients (\ref {eq:Gab})
and using the commutation relations of the generators of the Lorentz subgroup and also (\ref {eq:Fmmn}), (\ref {eq:Fmunu}),
we obtain a representation for the curvature tensor appro\-pri\-ate to known decomposition
of the Weyl--Cartan curvature tensor on symmetric and antisymmetric parts \cite {Sh-Str}:
\begin{equation}
\bar R^{a}\!_{b\mu\nu} = 2\partial_{[\mu}\Gamma^{a}\!_{\mid b\mid \nu ]} +
2\Gamma^{a}\!_{c [\mu} \Gamma^{c}\!_{\mid b\mid\nu ]}= -I\!_m\!^a\!_b F^{m}\!_{\mu\nu} +
\delta^{a}_{b}F_{\mu\nu}\; . \label{eq:Retr}
\end{equation}
\par
Using (\ref {eq:2217}) and taking into account (\ref {eq:Gab}) and (\ref {eq:Tmunu}), we obtain the expression for a
torsion tensor of spacetime manyfold:
\begin{equation}
T^{\lambda}\!_{\mu\nu} = 2\Gamma^{\lambda}\!_{[\nu\mu ]} = 2 h^\lambda\!_a h^{b}\!_{[\nu}
\Gamma^{a}\!_{\mid b \mid \mu]} + 2 h^\lambda\!_{a}\partial_{[\mu} h^{a}\!_{\nu ]} =
h^\lambda\!_a F^{a}\!_{\mu\nu}\;.\label{eq:Tor}
\end{equation}
\par
The relations (\ref {eq:2203}), (\ref {eq:Gab}), (\ref {eq:nem}), (\ref {eq:Retr}), (\ref {eq:Tor}) establish
the connection between the geometrical quantities of spacetime manifold and the relations of the gauge fields theory
for  the localized Poincar\'{e}--Weyl group.
\par
It is possible to obtain field equations by variation of the action integral for the general Lagrangian density
(\ref {eq:Ltot}) with respect to dynamical variables $\{A^m\!_\mu,\,A_\mu,\,h^a\!_\mu,\,\beta(x)\}$:
\begin{equation}
\frac{\delta{\cal L}}{\delta A^m\!_{\mu}} = 0\; , \qquad
\frac{\delta{\cal L}}{\delta A_{\mu}} = 0\; , \qquad
\frac{\delta{\cal L}}{\delta h^{a}\!_{\mu}} = 0\;, \qquad
\frac{\delta{\cal L}}{\delta \beta(x)} = 0\; .\label{eq:vurAmu}
\end{equation}
As it has been already pointed out, the last field equation is a consequence of the others. The first of these field equations
(\ref {eq:vurAmu}) can be represented as
\begin{eqnarray}
&&\partial_\nu \frac{\partial {\cal L}_0}{\partial F^{m}\!_{\mu\nu}} =
\frac{1}{2}\sqrt{\mid \bar g\mid}\, (S_{(0)m}^{\mu} + S^{\mu}\!_{m} )\; ,\label{eq:urFmn} \\
&& \sqrt{\mid \bar g\mid}\,S^{\mu}\!_{m} = - \frac{\partial {\cal L}_\psi}{\partial A^m\!_\mu} =
\frac{\partial{\cal L}_\psi}{\partial D_\mu \psi^A} I_m\!^A\!_B \psi^B\; ,\nonumber\\
&& \sqrt{\mid \bar g\mid}\,S_{(0)m}^{\mu} = - \frac{\partial{\cal L}_0}{\partial
A^m\!_\mu} =  2\frac{\partial{\cal L}_0}{\partial F^n\!_{\mu\nu}}
c_m\!^n\!_q A^q\!_\nu + 2\frac{\partial{\cal L}_0}{\partial F^c\!_{\mu\nu}}
I_m{}^c\!_a \,h^a\!_\nu \; .\label{eq:J(0)}
\end{eqnarray}
The second of these field equations (\ref {eq:vurAmu}) can be written down as follows
\begin{eqnarray}
&&\partial_\nu \frac{\partial {\cal L}_0}{\partial F_{\mu\nu}} =
\frac{1}{2}\sqrt{\mid \bar g\mid}\, (J_{(0)}^{\mu} + J^{\mu})\; ,\quad
\sqrt{\mid \bar g\mid}\,J^{\mu} = - \frac{\partial {\cal L}_\psi}{\partial A_\mu} =
\frac{\partial{\cal L}_\psi}{\partial D_\mu \psi^A} w \psi^A\; ,\label{eq:urFm} \\
&& \sqrt{\mid \bar g\mid}\,J_{(0)}^{\mu} = -\frac{\partial{\cal L}_0}{\partial
A_\mu} = - 2\frac{\partial{\cal L}_0}{\partial F^a\!_{\mu\nu}}h^a\!_\nu
- 8\frac{\partial{\cal L}_0}{\partial Q_{\mu}}\, .\label{eq:Jm}
\end{eqnarray}
Third of these field equations (\ref {eq:vurAmu}) can be represented as:
\begin{eqnarray}
&&\partial_{\nu} \frac{\partial{\cal L}_0}{\partial F^{a}\!_{\mu\nu}} =
-\frac{1}{2}\sqrt{\mid \bar g\mid}\, (t^{\mu}_{(0)a} + t^{\mu}_{(\psi)a} )\; , \label{eq:urTab} \\
&& \sqrt{\mid \bar g\mid}\,t^{\mu}_{(\psi)a} = \frac{\partial{\cal L}_\psi}{\partial h^{a}\!_{\mu}} =
h^\mu\!_a {\cal L}_\psi - \frac{\partial{\cal L}_\psi}
{\partial D_\mu \psi^{A}}D_a \psi^A\; , \label{eq:tQ} \\
&& \sqrt{\mid \bar g\mid}\,t^\mu_{(0)a} = \frac{\partial{\cal L}_0}{\partial h^{a}\!_{\mu}} =
h^\mu\!_a {\cal L}_0 - 2F^m\!_{a\nu}\frac{\partial{\cal L}_0}
{\partial F^m\!_{\mu\nu}} - 2F_{a\nu}\frac{\partial{\cal L}_0}{\partial F_{\mu\nu}}- \nonumber \\
&& - 2 F^b\!_{a\nu}\frac{\partial{\cal L}_0}{\partial F^b\!_{\mu\nu}} +
2A^m\!_\nu I_m{}^b\!_a \frac{\partial{\cal L}_0}{\partial F^{b}\!_{\mu\nu}}
- 2A_\nu \frac{\partial{\cal L}_0}{\partial F^{a}\!_{\mu\nu}} -
Q_a\frac{\partial{\cal L}_0}{\partial Q_{\mu}}\; . \label{eq:t0}
\end{eqnarray}
The quantity (\ref {eq:tQ}) represents the known expression for the canonical energy-momentum of an external
field \cite {He:rmf}.
\par
The quantities (\ref {eq:J(0)}), (\ref {eq:Jm}) and (\ref {eq:t0}) represent, accordingly, an internal spin momentum,
a proper dilatation current and an energy-momentum of free gauge fields. These currents are conserved in
the sum with the appropriate currents of an external field $\psi^A$. The given conservation laws are simple
consequences of the gauge field equations (\ref {eq:urFmn}), (\ref {eq:urFm}) and (\ref {eq:urTab}):
\begin{eqnarray*}
&&\partial_{\mu}\left (\sqrt{\mid \bar g\mid}\,(S_{(0)m}^{\mu} + S^{\mu}\!_{m})\right ) = 0\; ,
\quad \partial_{\mu}\left (\sqrt{\mid \bar g\mid}\,(J_{(0)}^{\mu} + J^{\mu})\right ) = 0\; ,\\
&& \partial_{\mu}\left (\sqrt{\mid \bar g\mid}\,(t_{(0)a}^{\mu} + t^{\mu}_{(\psi)a})\right ) = 0\; .
\end{eqnarray*}
\par
The field equations (\ref {eq:urFmn}), (\ref {eq:urFm}) and (\ref {eq:urTab}) can be represented in a geometrical form
with the help of a notion of the gauge derivative (\ref {eq:DmQ}), which however does not operate on the Greek coordinate
indexes:
\begin{eqnarray}
&&D_{\nu} \left (\frac{\partial{\cal L}_{0}}{\partial F^{m}\!_{\mu\nu}}
\right ) = \frac{1}{2}\sqrt{\mid \bar g\mid}\, S^{\mu}\!_{m} + \frac{\partial{\cal L}_{0}}
{\partial F^{b}\!_{\mu\nu}}I_m{}^b\!_a h^{a}\!_{\nu} \; ,\\
&&D_{\nu} \left (\frac{\partial{\cal L}_{0}}{\partial F_{\mu\nu}}\right ) =
\frac{1}{2}\sqrt{\mid \bar g\mid}\, J^{\mu} - \frac{\partial{\cal L}_{0}}{\partial F^{a}\!_{\mu\nu}}
h^{a}\!_{\nu} - 4\frac{\partial{\cal L}_0}{\partial Q_{\mu}}\; ,\\
&&D_\nu \left (\frac{\partial{\cal L}_{0}}{\partial F^{a}\!_{\mu\nu}} \right ) -
F^{m}\!_{a\nu}\frac{\partial{\cal L}_{0}}{\partial F^{m}\!_{\mu\nu}} -
F^b\!_{a\nu}\frac{\partial{\cal L}_{0}}{\partial F^{b}\!_{\mu\nu}} - \nonumber \\
&&- F_{a\nu}\frac{\partial{\cal L}_0}{\partial F_{\mu\nu}} - \frac{1}{2}Q_a
\frac{\partial{\cal L}_0}{\partial Q_{\mu}} + \frac{1}{2}h^{\mu}\!_a {\cal L}_{0} =
-\frac{1}{2}\sqrt{\mid \bar g\mid}\,t^{\mu}\!_{(\psi)a} \; .
\end{eqnarray}
These equations, the latter of which generalizes the Einstein equation to arbitrary Lag\-ran\-gi\-an, generalize
the field equations for the Poincar\'{e} group \cite{Fr:book}, \cite{FrIv}.
\par
By explicit calculations it is possible to prove the following theorem generalizing and specifying the theorem,
given in \cite {Fr:book} for the Poincar\'{e} group.
\begin {quote}
{\bf Theorem 4.} The gauge field equations (\ref {eq:urAm}) obtained according to the principle of stationary
action for a set of fields $\{A^m_a,\,A_a,\,A^k_a,\,\beta(x)\}$ are satisfied, if and only if the variational field
equations (\ref {eq:vurAmu}) are valid for a set of fields $\{A^m\!_\mu,\,A_\mu,\, h^a\!_\mu,\,\beta(x)) \}$ ,
provided that the tetrads $h^a\!_\mu$ are represented by the formulas (\ref {eq:h})--(\ref {eq:YY}).
\end {quote}
\par
The given theorem shows that in those cases, in which the tetrads structure is insigni\-fi\-cant, the various
choice of dynamic variables specified in the theorem 4, results in dynamically equivalent theories. However, in
some cases, for example, in a quantum theory of gravitational field, in which use of true gravitational
potentials is important, the given tetrad structure can appear essential. In this case the choice as
the dynamic variables the quantities $\{A^m_a,\,A_a,\,A^k_a,\,\beta(x)\}$ results in the theory,
richer by the opportunities.
\par
As analogy it is possible to point out a connection between metric and tetrad formula\-ti\-ons of a gravitational field
theory. The tetrad theory of gravitation, though results to the formulation equivalent for non-spinor matter to
the metric theory of gravitation with respect to field dynamical equations, nevertheless is richer by the opportunities
realized at the spinor description in Riemann geometry.

\section {Conclusion}
\setcounter {equation} {0}
\par On the basis of the general principles of the gauge fields theory the gauge theory for the Poincar\'{e}--Weyl group
was constructed. The abstract expression for the gauge derivative was obtained: $D_a = - A^{R}_{a} M_R$. It was
shown that, as against \cite {Char-Ta}, \cite {Kas}, the tetrads are not true gauge fields, but represent some
sufficiently complex functions from gauge fields: Lorenzian  $A^m_a $, translational $A^k_a $ and dilatational $A_a $,
the relation $A_a^k = D_a x^k $ being valid. It is possible to expect that the knowledge of the true gauge
potentials of a gravitational field appears essential at construction of the quantum theory of gravity.
\par
The gauge fields equations, which sources are an energy-momentum tensor, spin and orbital momenta, and also a dilatation
current of an external field are obtained. A new effect of the direct interaction of the Lorenzian gauge field with the
orbital momentum of an external field appears. A geometrical interpretation of the theory is developed and it is shown
that as a result of localization of the Poincar\'{e}--Weyl group spacetime becomes a Weyl--Cartan space. Also the
geometrical interpretation of the Dirac's scalar field $\beta$ \cite{Dir}) (and thereby of a dilaton field and also
of the Utiyama measure scalar field \cite{Uti}) as a component of the metric tensor of a tangent space in Weyl--Cartan
geometry was obtained. This field is necessary in construction a field theory in a Weyl--Cartan space \cite {Gr-Pap},
\cite{Nish}.
\par
The gauge invariant Lagrangian of the proper gauge fields is an arbitrary scalar function from the gauge strengthes
of the theory containing derivatives from the gauge fields not higher than the first order:
\[
{\cal L}_{0} = \sqrt{\mid \bar g\mid}\,L_{0}(F^{m}\!_{ab},\,F_{ab},\,F^{c}\!_{ab}, Q_{a},\,\beta (x))\;, \qquad
\sqrt{\mid \bar g\mid} = \beta^4\,h\;, \quad h = \det{(h^a\!_\mu)}\;.
\]
The most simple Lagrangian of such kind, allowing to all gauge fields to be realized dynamically, can be
 constructed as
\begin{eqnarray}
{\cal L}_{0} = 2hf_0 \beta^4(I_m\!^a\!_b F^m\!_a\!^b  + f F^m\!_{ab}F_m\!^{ab} + \rho F^c\!_{ab}F_c\!^{ab} +
 \lambda F_{ab}F^{ab}+ \xi Q_{a}Q^{a} + \zeta F^{c}\!_{ac}Q^{a} + \Lambda ) = \nonumber \\
2hf_0 \Bigg (\frac{1}{2}\beta^2 \bar R + 2f \bar R_{[ab]\mu\nu}\bar R^{[ab]\mu\nu} + 4\lambda (\partial_{[\mu} A_{\nu]})^2 +
\beta^2 (\rho T_{\lambda\mu\nu}T^{\lambda\mu\nu} + 64\xi A_{\mu}A^{\mu} + 8\zeta T^{\mu}A_{\mu}) - \nonumber\\
- 8(\zeta T^\mu + 16\xi A^\mu ) \beta \partial_\mu \beta + 64\xi g^{\mu\nu} ({\partial_\mu \beta}) ({\partial_\nu \beta})+
\Lambda \beta^4 \Bigg ) \; . \nonumber
\end{eqnarray}
Here $\bar R^a{}_{b\mu\nu}$ is the Weyl--Cartan curvature tensor, $\bar R$ is the Weyl--Cartan curvature scalar,
and the contractions of the Greek indexes are performed with the Riemann--Cartan metric tensor $g_{\mu\nu}$.
\par
The given Lagrangian has some distinctive features. First, it reproduces the quadratic Lagrangian of the
Poincar\'{e}-gauged theories of gravity \cite {Fr2}, \cite {He-CMN}, \cite {Fr:book}. Second, these Lagran\-gi\-an,
despite the gauge invariance, permits the presence of nonzero mass of the Weyl vector and therefore of the dilatation
gauge field, in contrast to \cite {Char-Ta}, \cite {Kas}. This circumstance means that the gauge field, introduced by
localization of the group of scale transformations, is not an electromagnetic field (as against initial idea of Weyl
and from \cite {Gr-Pap}), but a field of other nature, what was pointed to in \cite {Ut2}--\cite {Ut3}. A nonzero mass
of the Weyl field can play a positive role in interpretations of the modern observational data on the basis of using
post-Riemannian cosmological models \cite {Tuck}, \cite {CQG-03}, and also for a possible explanation of a graceful
exit from a stage of inflation. Besides, the last terms with a field $\beta (x)$ in this Lagrangian have the structure
of the Higgs Lagrangian \cite{Gr-Pap} and can play a determining role at spontaneous violation of scale invariance and
formation of mass of particles \cite {Fr-Sin}.
\newline
\par
Authors thank I. V. Tyutin for fruitful discussion of results of the work.

\end{document}